\def\beq{\begin{equation}}
\def\eeq{\end{equation}}
\def\bea{\begin{eqnarray}}
\def\eea{\end{eqnarray}}

\def\th{_{_{\rm th}}}

\documentstyle[aps,floats]{revtex}

\begin{document}

\renewcommand{\topfraction}{1.0}
\twocolumn[\hsize\textwidth\columnwidth\hsize\csname
@twocolumnfalse\endcsname

\title{Brans--Dicke wormholes endowed with torsion}
\author{Luis A. Anchordoqui
%\thanks{Electronic address: doqui@venus.fisica.unlp.edu.ar}
}

\address{Departamento de F\'{\i}sica, Universidad Nacional de La Plata\\
C.C. 67, 1900, La Plata, Buenos Aires, Argentina}
\maketitle

\begin{abstract}

A two--way traversable
wormhole solution in Brans--Dicke theory with torsion is obtained using the
method of massive thin shells.
The solution goes over general relativity for an infinite large value
of the coupling parameter, however, the Brans--Dicke scalar could never
be the ``carrier'' of exoticity that threads the wormhole throat.

\noindent{\it PACS numbers}: 04.20.Jb, 04.50.+h

\begin{center}
{\it To appear in Il Nuovo Cimento B}
\end{center}
\end{abstract}

\vskip2pc]

\newpage

Shortly after the paper of Morris and Thorne \cite{motho} the
fate of wormholes
has rapidly grown into an active area of research.
Intuitively speaking these wormholes are tunnels linking widely
separated regions of spacetime from where in-going causal curves can pass
through and become out-going on the other side.
The spherically symmetric line element for such a spacetime reads,
\begin{equation}
ds^2 = -\alpha^2 dt^2 + d\eta^2 + r^2 (d\theta^2 + \sin^2\theta d\phi^2)
\end{equation}
where the non-monotonic coordinate $r$ decreases from
$+\infty$ to a minimum value $r\th$,
representing the location of the throat of the wormhole and then
increases from $r\th$ to $+\infty$. The proper radial distance $\eta$ ranges
from $-\infty$ to $\infty$ with $\eta=0$ at the throat. The redshift function
$\alpha$ is positive everywhere, this ensures the absence of an event horizon.

It has long been known that when the gravitational field is
strong, the Brans Dicke (BD) scalar
may have some ``exotic'' effects on stellar configurations
through a
local modification of the gravitational constant by the matter
energy distribution \cite{kiril}.
In the context of traversable wormholes, Agnese and
La Camera showed that a static spherically symmetric vacuum BD
solution in the Jordan frame,
supports a two-way traversable wormhole for values of the BD
parameter $\omega < -2$ \cite{agnese}.
The domain of the coupling constant was
later extended to positive values \cite{nandi}.
Nevertheless, if one tries to get back general relativity (GR) these
wormholes are doomed. As direct consequence of the GR limiting process
the BD scalar must become a constant, triggering a gravitational collapse
of the throat. Moreover, the rigorous analysis performed
by Nandi et al. (in the conformally rescaled
Einstein frame) severely constrains the range of $\omega$ in the Jordan
frame, while at the same time shows that wormhole solutions
do not exist at all in the Einstein frame unless one is willing to
violate the energy conditions by choice \cite{nandi2}.

Whether or not
BD theory goes over GR
when the parameter $\omega$ has an infinitely large value is yet
to see the light of day. The study of the limits of spacetimes
depending on some parameter has been initiated by Geroch
\cite{geroch}, who called attention to the fact that the limit may
depend on the coordinate system chosen to perform the calculations.
More recently, based on the characterization of a given spacetime by
the Cartan scalars, it  was shown that the
BD solutions might not have a unique limit as  $\omega \rightarrow
\infty$ \cite{paiva}. As a matter of fact, an estimation of order
of magnitude
seems to indicate that BD theory always reduces to GR
for large values
of $\omega$ when the trace of
the matter stress energy tensor is not zero, however BD solutions do
not reduce to the corresponding GR limit in the trace-less case \cite{BS}.

Non-vacuum wormhole solutions (without
trace-less matter) in BD
theory were already reported
somewhere else
\cite{bdwh}. The ones
threaded with ``ordinary'' matter only exist in a very narrow
interval of the coupling parameter. Although the limiting process $\omega
\rightarrow \pm \infty$  does not destroy these wormholes, the matter
that threads their throats needs to be exotic.

The type of wormholes obtained by surgically grafting two identical
copies of various well known spacetimes (Schwarzschild \cite{S},
Reissner--Nordstr\"om \cite{RN}, Friedman--Robertson--Walker
\cite{FRW}, Schwarzschild--de Sitter \cite{SdS}) provides a particularly
elegant collection of exemplars which are not limited to be spherically
symmetric \cite{visser}. In this report, as
a natural extension of the Schwarzschild case, I
work out a traversable wormhole
solution in the modified BD theory with torsion
\cite{kim}. This particular solution, with non-zero stress energy on
the boundary layer between the two asymptotically vacuum flat regions, goes
over the Schwarzschild case for infinite large values of the coupling
parameter.

To construct the wormhole of interest, let me assume that the
matter which creates the gravitational field of the
wormhole is confined to a narrow region surrounding the throat $\eta
\in (-\epsilon, \epsilon)$.
Namely, consider a thin shell of stress energy, let
\begin{equation}
S_{\mu\nu} = \lim_{\epsilon \rightarrow 0} \int_{-\epsilon}^\epsilon
T_{\rho\sigma} \,h^\rho_ \mu \,h^\sigma _\mu \,d \eta
\end{equation}
denote the surface matter energy momentum tensor of such a shell;
where $T_{\mu\nu}$ stands for the matter stress energy
tensor and $h_{\mu\nu}$ projects general tensors down onto the subspace
spanned by the thin shell (remember that $\eta$ measures the
geodesic distance in the direction normal to the throat) \cite{israel}.
Associating negative values of $\eta$
to one side of the throat (say
lower universe) and positive values to the other side,
without
loss of generality the metric in the neighborhood of the shell can be
written as,
\begin{equation}
g_{\mu\nu} = \Theta (\eta) \,\,g^{+}_{\mu\nu} +
\Theta (-\eta)\,\, g^{-}_{\mu\nu}.
\end{equation}

Outside the shell the spacetime is described by the vacuum
solution
\begin{equation}
\alpha = \left(
1 - \frac{2 (2 \,\omega -1)}{(2\omega+3)} \frac{M}{r}\right)^{(2\omega+3)
/(4 \omega -2)},
\end{equation}
\begin{equation}
\frac{dr}{d\eta} = \sqrt{1 - \frac{ 2 \,(2 \omega -1) }{(2\omega+3)}
\frac{M}{r}}
\end{equation}
derived
from the modified Brans--Dicke action $S= \int d^4x \sqrt{-g}
(- \phi R + \omega \phi^{,\mu} \phi_{,\mu} / \phi)$ \cite{kimcho}. It is
worthy of notice that the differential spacetime manifold has a non-symmetric
affine connection with the torsion field being generated by the gradient
of the BD scalar $\phi$. The scalar curvature $R$ is that of $U_4$
theory, and $M$ is the
mass of the wormhole as measured by
distant observers. Note that I have previously used $\omega$ to
denote the usual BD
coupling parameter. The field equation of the torsion endowed BD
case are equivalent to those of vacuum by making $  \omega_{_{\rm
torsion}} \leftrightarrow
-(\omega_{_{\rm vacuum}} + 3/2)$;
see Eq. 25 of ref. \cite{kimcho} and Eq. 2.14 of ref. \cite{van}.
Before going on, and for the sake of completeness, let me remark that the
non-vanishing components of the torsion tensor are \cite{kimcho},
\begin{eqnarray}
F_{01}^0 = - F_{12}^2  = -F_{13}^3 & = & \frac{2M}{(2\omega+3) r^2}
\nonumber \\
 & \times & \left(
1 - \frac{ 2\,(2\omega-1)}{(2\omega+3)} \frac{M}{r} \right)^{-1/2}
\label{torsion}
\end{eqnarray}
and the BD scalar is given by,
\begin{equation}
\phi = \left( 1 - \frac{2\,(2\omega-1)}{(2\omega+3)} \frac{M}{r}
\right)^{-2/(2\omega-1)}.
\label{phi}
\end{equation}
It follows from Eqs. (\ref{torsion}) and (\ref{phi}) that in the limit
$\omega \rightarrow \infty$ the torsion tensor
vanishes, and the BD scalar approaches to 1, the GR value.

Hereafter, I shall assume symmetry
under interchange of asymptotically flat regions $\pm \leftrightarrow
\mp$. However, one should note that
this requirement is not essential to the definition of a
traversable wormhole. Actually, if the wormhole throat is taken to
have different masses (or equivalently the behaviour of the BD scalar is
different) in the upper and lower universe, the wormhole will be
non-symmetric \cite{frono}.

Since the stress energy tensor is of delta function type at the
boundary of the two regions, $S_{\mu\nu}$ can be expressed in terms
of the jump of the second fundamental forms
\begin{equation}
S^\mu_\nu = ({\cal K}^\mu_\nu - \delta^\mu_\nu \,{\cal K})
\end{equation}
where ${\cal K}^\mu_\nu = K^{\mu\,+}_\nu - K^{\mu\,-}_\nu$, being
$K^{\mu\,\pm}_\nu$ the second fundamental forms evaluated above and
below the thin shell,
\begin{equation}
K^{\mu\,\pm}_\nu = \left. \frac{1}{2} g^{\mu\nu}
\frac{dg_{\mu\nu}}{d\eta} \right|_{\eta \rightarrow \pm 0}.
\end{equation}
The surface stress energy has to satisfy a condition of pressure
balance together with a constraint reflecting the fact that stress
energy may be exchanged between the layers. For the static,
spherically, and reflection symmetric
cases under consideration these constraints are automatically
satisfied.
Concerning the role of the BD scalar on the layer, there exists some
regions of the parameter space in which the scalar field behaves like a
domain wall (the reader is
referred to Section III part B of ref. \cite{letwang}).
However, for hypothetical wormhole's engineering considerations, one
has nothing to show for this repulsive gravitational field. This
could be checked by computing the surface matter energy density
\begin{equation}
S_t^t = -\frac{\phi\th}{2\pi r\th} \sqrt{1 -
\frac{2 \,(2 \omega -1)}{(2\omega+3)}\frac{M}{r\th}}
\label{1}
\end{equation}
and the surface matter tension
\begin{equation}
S_\theta^\theta = S_\phi^\phi= -\frac{\phi\th}{4 \pi
r\th} \left[ \frac{ 1- (2\omega -
5)\,M/(2\omega+3)\,r\th}{\sqrt{1-2\,(2\omega-1)\,M/(2\omega+3)\, r\th}}
\right].
\label{2}
\end{equation}
It follows from Eqs. (\ref{1}) and (\ref{2}) that one can not
keep the
surface matter stress energy from violating the energy conditions.
It is important to stress
that the expectation values of the stress
energy tensor with respect to certain quantum states are known to
violate these conditions \cite{qft}. On a purely classical side,
it has been recently suggested that some of
the gamma ray bursts detected by the {\it Burst and Transient Source
Experiment} (BATSE) might be the telltale signature of gravitational negative
anomalous compact objects \cite{grbwh}.

It is straightforward to prove that the solution will survive a
GR limiting process. Moreover, in such a limit the classical wormhole
solution
introduced by Visser and its equation of state are regained \cite{S}.

Summing up, using the massive thin shell formalism I found out a
wormhole solution of BD theory. The matter at the throat is
necessarily exotic. The solution is well behaved for every value
of the coupling constant, and in the limit
$\omega \rightarrow \pm \infty$ one
recovers GR which is believed to be the best theory of gravitation.

\vspace{1cm}

This work has been partially supported by FOMEC.

\end{document}